\documentclass[12pt]{article}
\usepackage{cite}
\usepackage{textcomp}
\usepackage{amsmath,amssymb,amsfonts}
\usepackage{algorithmic}
\usepackage{graphicx}
\usepackage{textcomp}
\usepackage{xcolor}\hyphenpenalty=50000

\begin{document}
\mbox{}\\
{{\huge\bf{Ultrasound homogenises\\suspensions of hydrophobic\\particles}}}\\
\sloppy

Michiel Postema$^{1}$\footnote[2]{Correspondence to: Michiel.Postema@wits.ac.za}, Ryonosuke Matsumoto$^2$, Ri-ichiro Shimizu$^2$, Albert~T.~Poortinga$^3$, and Nobuki Kudo$^2$ 

($^1$School Elec. \& Inform. Eng., Univ. Witwatersrand, Johannesburg, South Africa; $^2$Grad. School Inform. Sci. \& Technol., Hokkaido Univ., Sapporo, Japan; $^3$Dept. Mech. Eng., Eindhoven Univ. Technol., Eindhoven, Netherlands)\\

{\it Submitted to the 40th Symposium on UltraSonic Electronics (USE2019).}

\section{Introduction}
Ultrasound baths are commonly used for cleaning metal objects.$^1$ But they have also been proposed as convenient replacement of acoustic shakers for bulk mixing.$^2$ These latter hand-held devices mix small quantities whilst requiring the ultrasonic tip to be dipped in the liquid to be mixed, which may be undesirable.$^3$ 

Hydrophobic particles inherently resist being suspended. Hydrophobic particles might be regarded as tiny solid particles surrounded by a thin gaseous shell.$^4$ It has been hypothesised that hydrophobic particles act as cavitation nuclei.$^4$ This cavitation behaviour would explain the translation speeds observed when hydrophobic polystyrene microspheres were driven through a liquid medium by means of ultrasound.5 These translation speeds corresponded to those observed with gas microbubbles of similar sizes.$^5$

If hydrophobic particles do have a thin gaseous layer surrounding the solid cores, a sound field of sufficient pressure amplitude might force the gas layer to form and inertial cavity and subsequently fragment during the collapse phase.$^6$
In this study, we investigated whether hydrophobic particles can be forced to suspend by using ultrasound.

\section{Materials and Methods}

The first material studied was TIMCAL SUPER C65 Carbon Black (EQ-Lib-SuperC65) conductive additive (MTI Corportation, Richmond, CA, USA), hereafter referred to as C65. This material has shown to generate harmonic response to a low-amplitude, i.e., non-destructive, ultrasound field.$^7$

The second material studied was hydrophobically modified Zano 10 Plus zinc oxide (Umicore, Brussel, Belgium), hereafter referred to as ZnO. This material is the core component inside acoustically active antibubbles.$^8$

For each material, samples of 1 mg were deposited into a FALCON\textsuperscript{\textregistered} 15 mL High-Clarity Polypropylene Conical Tube (Corning Science México S.A. de C.V., Reynosa, Tamaulipas, Mexico), after which 5 mL of 049-16787 Distilled 
Water (FUJIFILM Wako Pure Chemical Corporation, Chuo-Ku, Osaka, Japan) was added.
Each tube was individually held by hand for 1 minute in a 2510J-MT BRANSONIC\textsuperscript{\textregistered} ULTRASONIC CLEANER (BRANSON ULTRASONICS CORPORATION, Danbury, CT, USA) filled with degassed water. The ultrasonic cleaner operated at a transmitting frequency of 45 kHz.

Following the ultrasonic treatment, for each material, 200 $\mu$L was pipetted into the observation chamber of a high-speed observation system.9 The observation chamber was placed under an IX70 microscope (Olympus Corporation, Shinjuku-ku, Tokyo, Japan) with a LUMPlan FI/IR 40$\times$ (N.A. 0.8) objective lens. Attached to the microscope was an HPV-X2 high-speed camera (Shimadzu, Nakagyo-ku, Kyoto, Japan), operating at 10 million frames per second.$^{10}$ During camera recording, the materials were subjected to ultrasound pulses, each consisting of 3 cycles with a centre transmitting frequency of 1 MHz and a peak-negative pressure of 200 kPa, from a laboratory-assembled single-element transducer.$^{9,10}$ The signal fed into the transducer was generated by an AFG320 arbitrary function generator (Sony-Tektronix, Shinagawa-ku, Tokyo, Japan) and amplified by a UOD-WB-1000 wide-band power amplifier (TOKIN Corporation, Shiroishi, Miyagi, Japan).

\section{Results and Discussion}

Photographs of a test tube containing C65 before and after sonication in the ultrasonic cleaner are shown in \textbf{Fig. 1}. The carbon black has been mixed optically homogeneously in the water.
\begin{figure}[htbp]
\centerline{\includegraphics[width=0.5\linewidth]{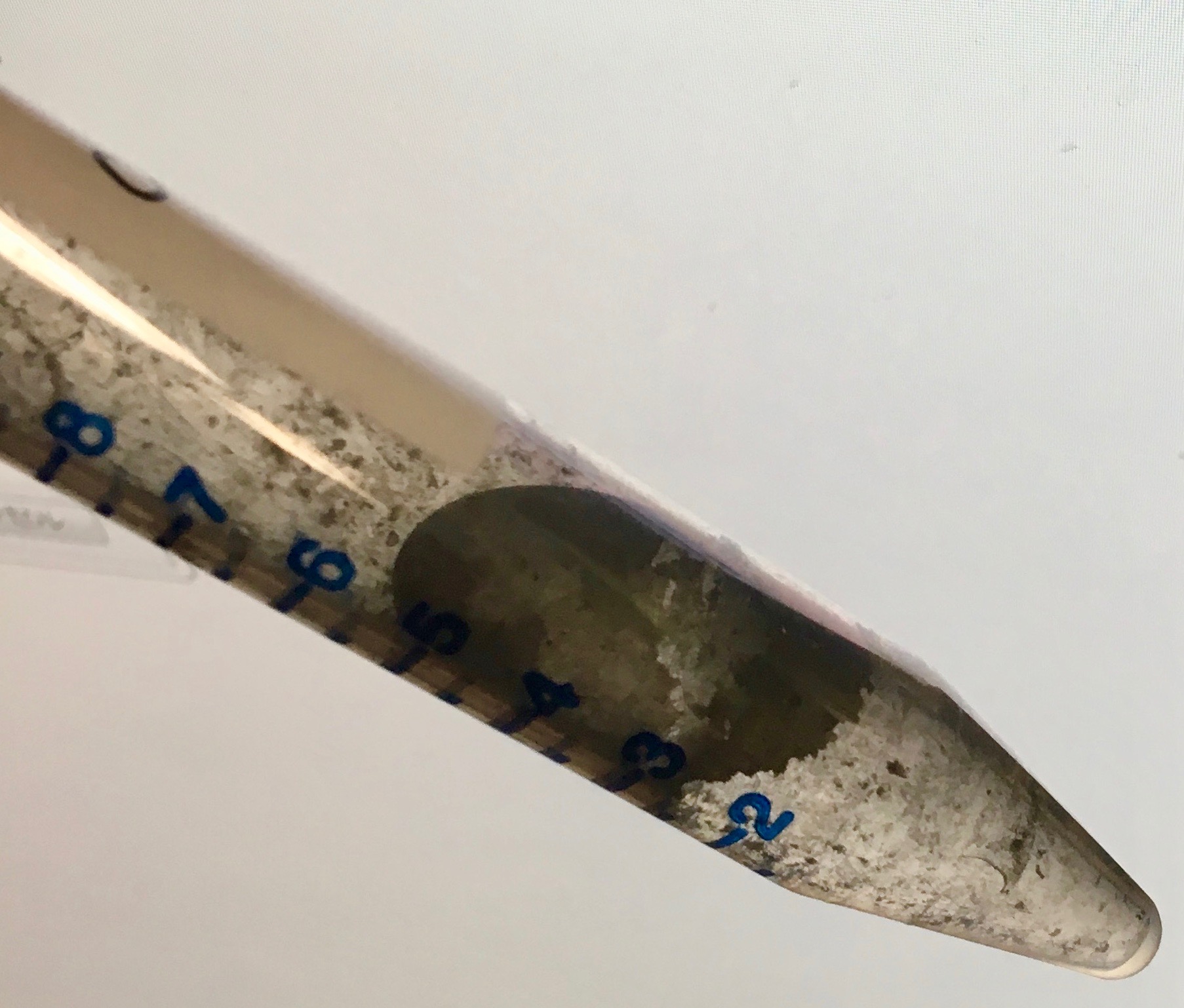}
\includegraphics[width=0.27\linewidth]{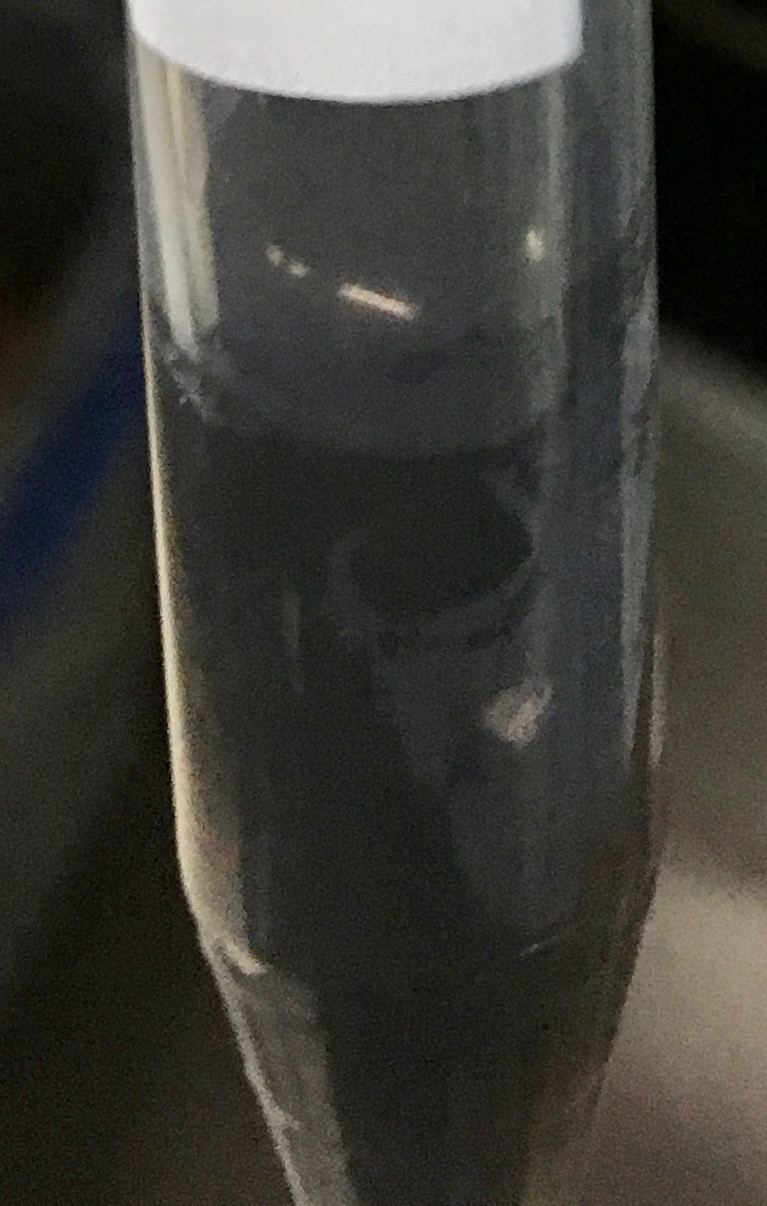}
}
\caption{C65 before (left) and after (right) 1-minute sonication in the ultrasonic cleaner.} 
\end{figure}

Photographs of a test tube containing zinc oxide before and after sonication in the ultrasonic cleaner are shown in \textbf{Fig. 2}. The white ZnO has been mixed optically homogeneously in the water.
\begin{figure}[htbp]
\centerline{\includegraphics[width=0.452\linewidth]{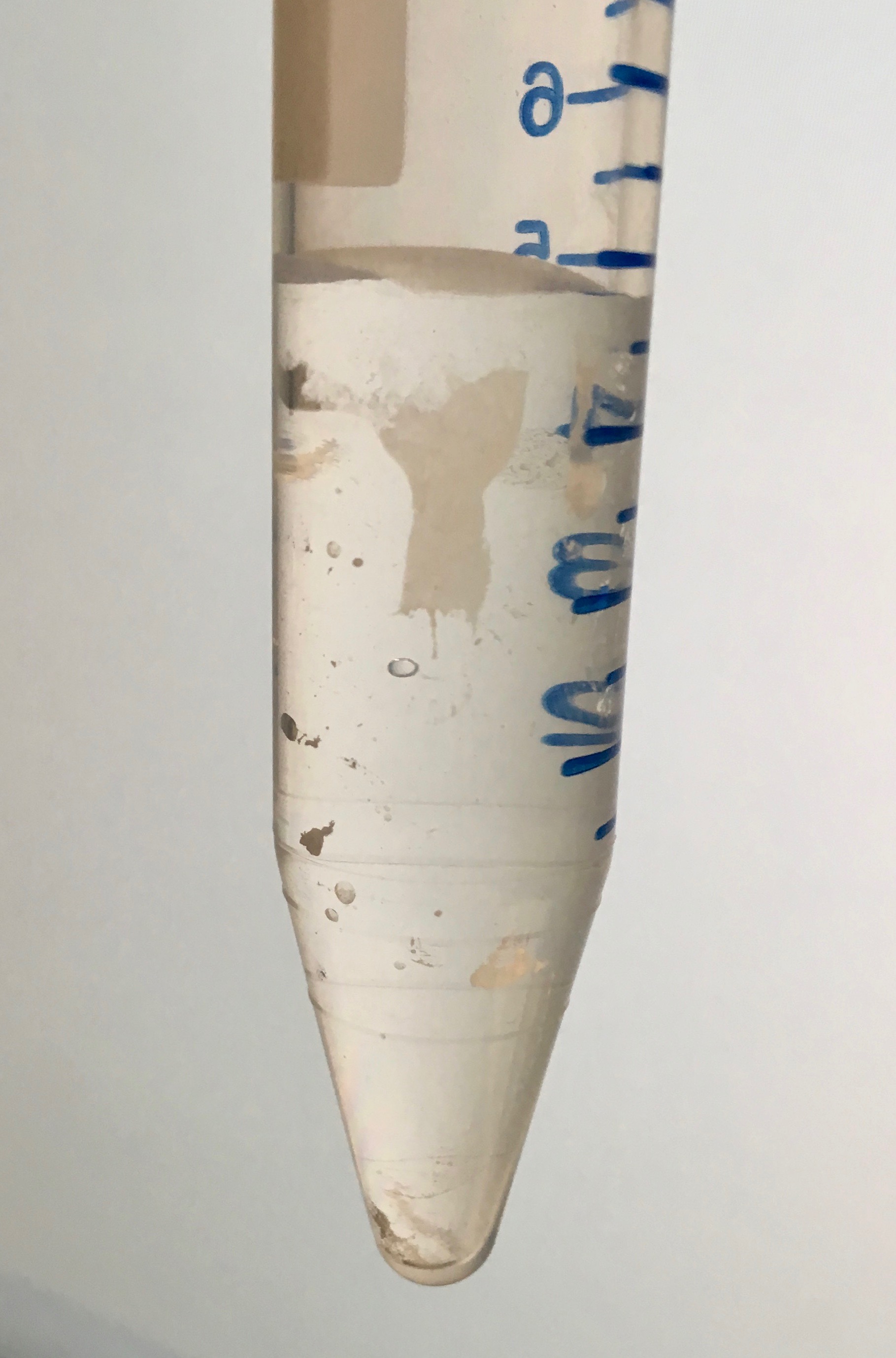}
\includegraphics[width=0.505\linewidth]{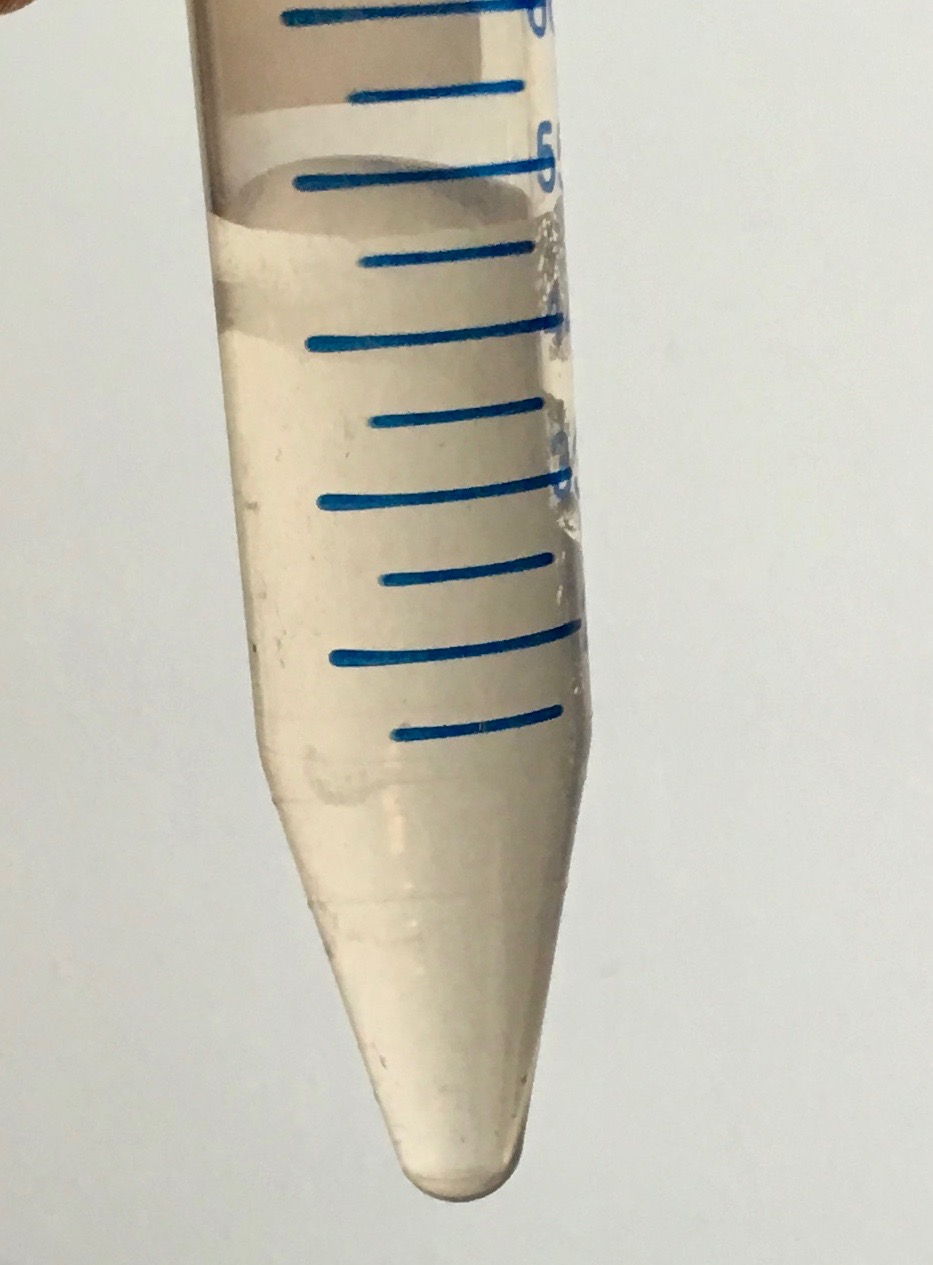}
}
\caption{ZnO before (left) and after (right) 1-minute sonication in the ultrasonic cleaner.} 
\end{figure}

\textbf{Fig. 3} shows three frames of a representative example from 56 high-speed video recordings of C65 and ZnO, each consisting of 256 frames. The frames shows a hydrophobic C65 particle before, during, and after 1-MHz sonication. During sonication, an inertial cavity can be clearly appreciated. After sonication, the cavity is seen to be detached and located to the upper right of the deformed or fragmented particle. In later frames (not shown), the cavity has dissolved, leaving the deformed or fragmented C65 particle. During subsequent ultrasound pulses, the C65 particle remains unaffected (not shown).

\begin{figure}[htbp]
{\includegraphics[width=0.65\linewidth]{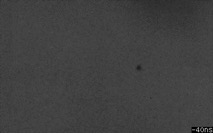}
\includegraphics[width=0.65\linewidth]{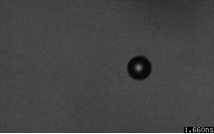}
\includegraphics[width=0.65\linewidth]{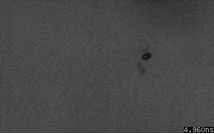}}
\caption{C65 acting as a cavitation nucleus, before (top), during (middle), and after (bottom) sonication. Each frame width corresponds to 145 $\mu$m. Time stamps are shown in the lower right corner of each frame.}
\end{figure}

In general for all experiments, during the first pulse, the hydrophobic particles of both materials were observed to grow into inertial cavities, forcing the gas layer surrounding the particle to be released and dissolved. After the first pulse, no further cavitation activity was observed. This indicates that sonication can remove the thin gas layer surrounding a hydrophobic particle.

\section{Conclusions}
Hydrophobic particles of the materials C65 and ZnO can be forced to be suspended in water using ultrasound. The high-speed observations confirm that hydrophobic particles can act as cavitation nuclei. The lack of cavitation after the first pulse indicates that the gas layer surrounding the hydrophobic particle dissolves after inertial cavitation.

\section*{References}
\begin{enumerate}
\item	P. Gehrke, R. Smeets, M. Gosau, et al.: In Vivo 33 (2019) 689.
\item	C. Shi, W. Yang, J. Chen, et al.: Ultrason. Sonochem. 37 (2017) 648.
\item	Y. Qiu, H. Wang, C.E.M. Demore, et al.: Sensors 14 (2014) 14806.
\item	P. Attard: Adv. Colloid Interface Sci. 104 (2003) 75.
\item	N. Mazzawi, M. Postema and E. Kimmel: Acta Phys. Pol. A 127 (2015) 103.
\item	M. Postema and G. Schmitz: Ultrason. Sonochem. 14 (2007) 438.
\item	M. Postema, S. Phadke, A. Novell, et al.: Proc. IEEE AFRICON (2019) submitted.
\item	M. Postema, A. Novell, C. Sennoga, et al.: Appl. Acoust. 137 (2018) 148.
\item N. Kudo: IEEE Trans. Ultrason. Ferroelect. Freq. Control 64 (2017) 273.
\item S. Imai, N. Kudo: Proc. IEEE Ultrason. Symp. (2018) 8579713.
\end{enumerate}
\end{document}